\documentclass[Physsubmission, Phys]{SciPost}

\hypersetup{
    colorlinks,
    linkcolor={red!50!black},
    citecolor={blue!50!black},
    urlcolor={blue!80!black}
}

\usepackage[bitstream-charter]{mathdesign}
\usepackage{todonotes}
\DeclareSymbolFont{usualmathcal}{OMS}{cmsy}{m}{n}
\DeclareSymbolFontAlphabet{\mathcal}{usualmathcal}

\begin{document}

\begin{center}{\Large \textbf{
Precision measurements on dipole moments of the tau and hadronic multi-body final states
}}\end{center}

\begin{center}
F.~Krinner\textsuperscript{1*},
S.~Paul\textsuperscript{1,2}
\end{center}

\begin{center}
{\bf 1} Max-Planck-Institut f\"ur Physik, M\"unchen, Germany
\\
{\bf 2} Technische Universit\"at M\"unchen, Physik-Department, E18, M\"unchen, Germany
\\
* fkrinner@mpp.mpg.de
\end{center}

\begin{center}
\today
\end{center}


\definecolor{palegray}{gray}{0.95}
\begin{center}
\colorbox{palegray}{
  \begin{minipage}{0.95\textwidth}
    \begin{center}
    {\it  16th International Workshop on Tau Lepton Physics (TAU2021),}\\
    {\it September 27 – October 1, 2021} \\
    \doi{10.21468/SciPostPhysProc.?}\\
    \end{center}
  \end{minipage}
}
\end{center}

\section*{Abstract}
{\bf
In the framework of precision experiments, the search for electric dipole moments and the precise determination of magnetic dipole moments (g-2)
have since long been of prime interest.
Hadronic decays offer the best accuracy, since only the kinematic information carried by a single neutrino per decay is lost.
Thus, they reveal more easily precious information on the helicity of the initial tau lepton. However, in contrast 
to one- or two-body hadronic final states, the description of hadronic multi-body final states depends on the model for the 
hadronic current. In this work, we determine how 
the choice of a hadronic model impacts the extraction of tau electric and magnetic dipole moments.
}


\section{Introduction}
\label{sec:intro}
In light of the recent result on the 
anomalous magnetic moment of the muon $(g-2)_\mu$ \cite{fermilab},
the study of the magnetic moments $\mu_\tau$ of the tau lepton
receives new attention motivated by the mass of the tau lepton 
being about 17 times larger than the mass of the muon.
In addition, electric dipole moments like $d_\tau$ 
are a key observable to search for effects of new physics,
as well.

Both $\mu_\tau$ and $d_\tau$ may be studied 
measuring to high precision 
the production and subsequent decays of $\tau^\pm$-pairs in 
$e^+$-$e^-$-collisions at $B$-factories.
Since the tau lepton has many different decay modes with none of them being dominant, 
the inclusion of the largest number of decay channels 
is required to statistically improve 
the precision of such measurements.

For most of the dominating decay modes like $(\pi\nu)$ or $(\ell\nu_\ell\nu_\tau)$,
we can construct the decay amplitudes from first principles. However, the amplitudes for 
hadronic multi-body final states depend on modelling the hadronic systems.
Hadronic decays are particularly suited since they include only a single 
escaping neutrino in contrast to leptonic decays with two neutrinos ($\nu_\tau$ and $\nu_\ell$) 
missing in the final state. The latter results in large uncertainties in the 
reconstruction of the total event kinematics. The inclusion of hadronic decays 
(37\% branching fraction), however, requires their very good understanding in 
order to reduce systematic uncertainties connected to their modelling. 
This is particularly true for hadronic multi-body ($n>2$) decays, which 
make up about $40\%$ of all hadronic decays \cite{PDG}.
Since for the measurement of the electric and 
magnetic moments the full $\tau^\pm$-pair 
event is studied, the inclusion of multi-body final states 
improves the exploitation of available data sets, presently mostly constrained to
final states of $(e^\pm\nu_e\nu_\tau)$, $(\mu^\pm\nu_\mu\nu_\tau)$, 
$(\pi^\pm\nu_\tau)$ and $(\rho^\pm\nu_\tau)$, commonly used for such measurements \cite{belle}.

The choice of the model for hadronic multi-body final states
is not unique and we must thus estimate the impact of the differences between the 
true model and the analysis model on the measurement of the tauon 
electric and magnetic moments. 

This article is structured as follows: in Sec.~\ref{sec:formfactors}
we introduce the form factors $F_2$ and $F_3$ and construct the 
spin-density matrix for the production of $\tau^\pm$-pairs. In Sec.~\ref{sec:neutrino},
we elaborate on the effects of the escaping neutrinos on the determination of 
$F_{2/3}$. 
In Sec.~\ref{sec:hadroniccurrentmodel}, 
we introduce the hadronic model required for hadronic multi-body final states.
In Sec.~\ref{sec:optimalobservables}, we construct so-called optimal observables used to extract the 
value of $F_{2/3}$ from data and use them to study the impact of the hadronic model on the 
measurement of $F_{2/3}$, as described in Sec.~\ref{sec:studies} using simulated data.

\section{Form factors}
\label{sec:formfactors}
\begin{figure}[h]
\centering
\includegraphics[width=0.45\textwidth]{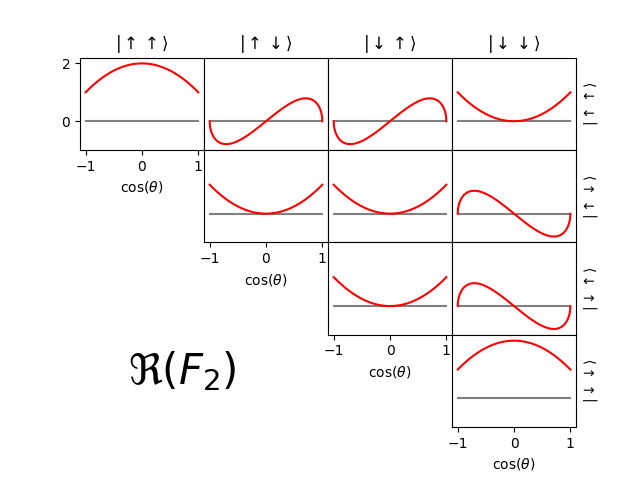}
\includegraphics[width=0.45\textwidth]{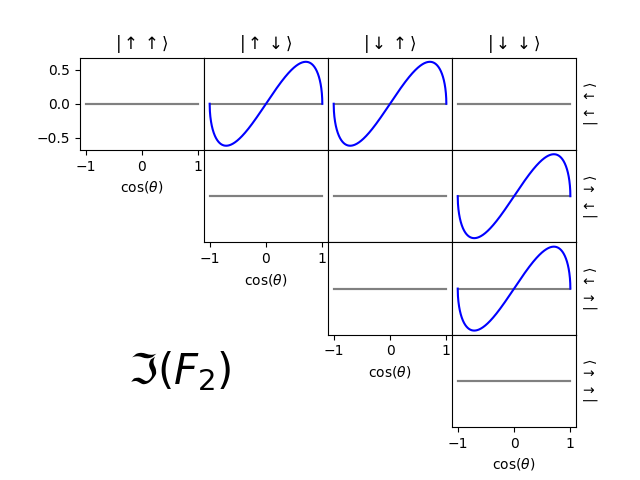}\\
\includegraphics[width=0.45\textwidth]{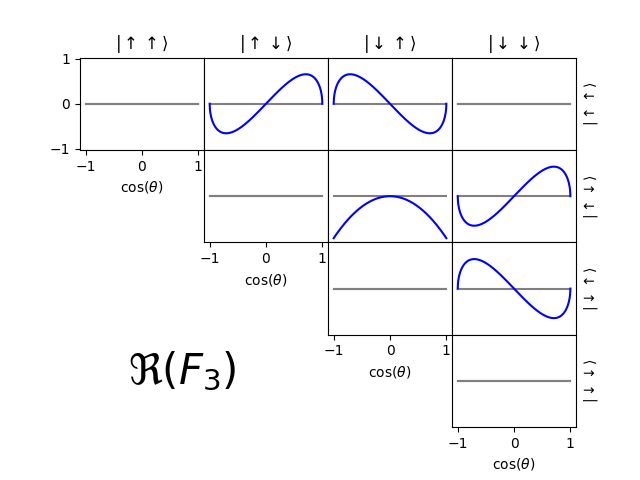}
\includegraphics[width=0.45\textwidth]{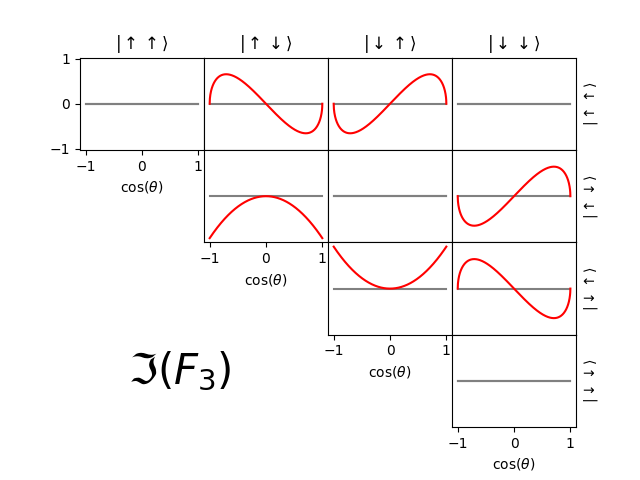}
\caption{Contributions from the form-factors $F_{2/3}$ to the $\tau^\pm$-pair production spin-density 
         matrix as function of the production angle $\cos(\theta)$.
         Contributions from the real and imaginary parts of $F_{2/3}$ are on the left and right, respectively.
         The influence of $F_2$ and $F_3$ are shown on the top and bottom row. Real and imaginary parts of the spin-density matrix
         are shown as red and blue lines, respectively.
         The vertical axis range is the same for all 10 plots of one contribution and is indicated on the leftmost
         sub-plot. Entries below the diagonal are omitted, since they are hermitian conjugates of the 
         upper-diagonal entries.}
\label{fig:SDM}
\end{figure}

The coupling of $\tau^\pm$-pairs to the photon field 
is described by:
\begin{equation}
-e \bar u_{\lambda_-} \Gamma^\mu v_{\lambda_+}
,\end{equation}
where $u_{\lambda^-}$ and $v_{\lambda^+}$ are the usual Dirac-spinors 
of the tauons with helicities $\lambda_\pm$ and the $\Gamma^\mu$ is given by:
\begin{equation}\label{eq:formfactors}
\Gamma^\mu = F_1(q^2) \gamma^\mu + \frac{iF_2(q^2)}{2m_\tau} \sigma^{\mu\nu} q_\nu + \frac{F_3(q^2)}{2m_\tau} \sigma^{\mu\nu} \gamma^5 q_\nu
,\end{equation}
where $q^\mu$ is the total four-momentum. $F_1(q^2)$ is the Dirac form-factor and $F_2(q^2)$ is the Pauli form-factor.
$F_{2/3}$ are connected to the electric and magnetic dipole moments via:
\begin{equation}
F_2(q^2=0)+1 = \frac{2m_\tau}{eQ_\tau} \mu_\tau \quad\text{and}\quad F_3(q^2=0) = \frac{2m_\tau}{eQ_\tau} d_\tau
.\end{equation}
The amplitude for the $\tau^\pm$-pair production in $e^+$-$e^-$-collisions is then 
given by:
\begin{equation}
\mathcal{A}_{\lambda_{e^-}\lambda_{e^+}\lambda_-\lambda_+} = \frac{e^2}{q^2}\cdot\bar v_{\lambda_{e^p}} \gamma_\mu u_{\lambda_{e^-}}\cdot u_{\lambda_-} \Gamma^\mu v_{\lambda_+}
,\end{equation}
where $\lambda_{e^-}$ and $\lambda_{e^+}$ are the helicities of the beam particles.
From this amplitude, we can construct the spin-density-matrix for the $\tau^\pm$-pair, 
which for the case of unpolarized $e^+$ and $e^-$ beams is given by:
\begin{equation}\label{eq:SDMconstruction}
\chi_{\lambda_-\lambda_+\lambda_-^\prime\lambda_+^\prime} = \frac14\sum_{\lambda_{e^\pm}} \mathcal{A}_{\lambda_{e^-}\lambda_{e^+}\lambda_-\lambda_+}^* \mathcal{A}_{\lambda_{e^-}\lambda_{e^+}\lambda_-^\prime\lambda_+^\prime}
.\end{equation}
Non-zero values of the form factors $F_{2/3}$ change the 
spin-density matrix and thus the spin-correlations of the produced $\tau^\pm$-pair. 
The changes to the spin-density matrix elements related to $\Re/\Im(F_{2/3})$ are shown 
as function of the $\cos(\theta)$ in Fig.~\ref{fig:SDM}, where $\theta$ is the production 
angle of the $\tau^-$ with respect to the incoming electron. The varying symmetry properties 
of the spin density matrix elements can be seen, and only $\Re(F_2)$ changes the total 
production cross-section\footnote{Comparing the spin-density matrix contributions to the ones
given in Ref.~\cite{NLO}, we find similarities between the contributions from $\mathcal{O}(\alpha^3)$ and 
$F_2$, resulting in the bias observed in Ref.~\cite{NLO}.}. For most form factors and spin combinations, 
extreme forward and backward angles as well as 90 degrees provide no sensitivity. Here, production angles 
around  $\pm45$ degrees seem most important. 

Since tauons decay before crossing any detector element, spin-correlations of the $\tau^\pm$-pair can only be accessed 
through the angular distributions of the $\tau^\pm$ decay products. In this work, we focus on 
such spin correlations in $\tau^\pm$-pair production\footnote{In this process, the kinematic 
range for the measurement of
$F_{2/3}(q^2)$ is limited to $q^2 > 4m_\tau^2$.} and the corresponding intensity distribution 
$\mathcal{I}$ of the decay products of both $\tau^\pm$ is constructed via:
\begin{equation}\label{eq:intens}
\mathcal{I} = \sum_{\lambda^{(\prime)}_\pm} \chi_{\lambda_-\lambda_+\lambda_-^\prime\lambda_+^\prime}\cdot D^-_{\lambda_-\lambda_-^\prime}\cdot D^+_{\lambda_+\lambda_+^\prime}
,\end{equation}
where $D^\pm_{\lambda_\pm\lambda_{pm}^\prime}$ are the 
spin-density matrices for the $\tau^\pm$ decays.

\section{Effects of neutrino kinematics}
\label{sec:neutrino}
In principle, all decay modes of the $\tau$-lepton are suitable for the 
determination of the form factors $F_{2/3}$. Simple accuracy studies 
similar to studies presented in Sec.~\ref{sec:studies} show that the 
accuracy for the form-factors $F_{2/3}$ is similar for all combinations 
of the dominant $\tau^\pm$ decay modes. This, however, requires the final-state
kinematic information to be complete, and thus the intensity distribution $\mathcal{I}$ given in 
Eq.~(\ref{eq:intens}) can simply be calculated.

However, since in every decay at least one neutrino is escaping, 
calculating the intensity distribution is no longer possible and unmeasurable degrees 
of freedom have to be integrated out. In events, where only a single neutrino escapes 
in each tau decay---making two in total---a two-fold kinematic ambiguity arises
for the direction of the tauons that has to 
be averaged in the calculation of $\mathcal{I}$. For this, both $\tau^-$ and 
$\tau^+$ must decay hadronically. For every $\tau^\pm$ decaying leptonically, an additional integration 
has to be performed:
\begin{equation}\label{eq:integ}
\mathcal{I} \to \iiint\mathcal{I}\,\mathrm{d}\phi\,\mathrm{d}\!\cos\theta\,\mathrm{d}m_{\nu\bar\nu}^2
,\end{equation}

where $m_{\nu\bar\nu}$ is the 
invariant mass of the escaping $(\nu\bar\nu)$-system and $\theta$ and $\phi$
the polar and azimuthal angle of the $\tau$ neutrino within this system.

This loss of kinematic information decreases the accuracy for the 
form factors $F_{2/3}$ depending on the particular combination of decay channels used.
This reduction in accuracy is summarized in Table~\ref{tab:resolution}, comparing the 
accuracies $\delta$ obtained with integrated and with fully known kinematic information:
\begin{equation}
x_{\delta\Re/\Im(F_{2/3})} = \frac{\delta_\text{integrated}\Re/\Im(F_{2/3})}{\delta_\text{known}\Re/\Im(F_{2/3})}
.\end{equation}
The averaging of the two-fold ambiguity for hadronic decays thus leads to a small decrease 
in accuracy, while the integration given in Eq.~(\ref{eq:integ}) for leptonic decays 
has a much larger effect, in particular for $\Re(F_3)$.

Thus, an increase of the usable data set of hadronic decays would 
improve the accuracies for $F_{2/3}$. In this work, we discuss the inclusion of the 
multi-body final-state with the highest branching fraction of $9.31\%$\cite{PDG}:
$\tau^\pm\to\pi^\mp\pi^\pm\pi^\pm + \nu$. Since this decay mode can be 
combined with all available decay modes of the opposite-sign $\tau$, its inclusion 
would increase the number of available purely hadronic events by a factor 1.57.

\begin{table}\begin{center}
 \caption{Decrease of accuracy due to the loss of kinematic information due to escaping
          neutrino kinematics for
          16 combinations of $\tau^\pm$ decay modes. The numbers are based on sets of $10^6$ simulated events.}
     \begin{tabular}{l|l|r|r|r|r}\label{tab:resolution}
    $\tau^-$ mode & $\tau^+$ mode & $x_{\delta\Re(F_2)}$  & $x_{\delta\Im(F_2)}$ & $x_{\delta\Re(F_3)}$ & $x_{\delta\Im(F_3)}$ \\\hline
    $\pi^-\nu_\tau$  & $\pi^+\bar\nu_\tau$                                              &$1.09$ & $1.60$ & $1.61$  & $1.06$ \\
    $\pi^-\nu_\tau$  & $\rho^+\bar\nu_\tau$                                             &$1.11$ & $1.19$ & $1.19$  & $1.10$ \\ 
    $\pi^-\nu_\tau$  & $\mathrm{e}^+\bar\nu_\tau\nu_\mathrm{e}$                         &$2.07$ & $1.75$ & $3.84$  & $1.92$ \\ 
    $\pi^-\nu_\tau$  & $\mu^+\bar\nu_\tau\nu_\mu$                                       &$2.06$ & $1.72$ & $3.73$  & $1.92$ \\ 
    $\rho^-\nu_\tau$ & $\pi^+\bar\nu_\tau$                                              &$1.11$ & $1.19$ & $1.19$  & $1.10$ \\
    $\rho^-\nu_\tau$ & $\rho^+\bar\nu_\tau$                                             &$1.11$ & $1.26$ & $1.15$  & $1.10$ \\
    $\rho^-\nu_\tau$ & $\mathrm{e}^+\bar\nu_\tau\nu_\mathrm{e}$                         &$2.03$ & $1.79$ & $3.18$  & $1.92$ \\
    $\rho^-\nu_\tau$ & $\mu^+\bar\nu_\tau\nu_\mu$                                       &$2.04$ & $1.79$ & $3.17$  & $1.92$ \\
    $\mathrm{e}^-\nu_\tau\bar\nu_\mathrm{e}$ & $\pi^+\bar\nu_\tau$                      &$2.09$ & $1.81$ & $3.97$  & $2.04$ \\
    $\mathrm{e}^-\nu_\tau\bar\nu_\mathrm{e}$ & $\rho^+\bar\nu_\tau$                     &$2.03$ & $1.75$ & $3.32$  & $1.82$ \\
    $\mathrm{e}^-\nu_\tau\bar\nu_\mathrm{e}$ & $\mathrm{e}^+\bar\nu_\tau\nu_\mathrm{e}$ &$3.45$ & $2.28$ & $21.83$ & $3.33$ \\
    $\mathrm{e}^-\nu_\tau\bar\nu_\mathrm{e}$ &$\mu^+\bar\nu_\tau\nu_\mu$                &$3.73$ & $2.28$ & $11.72$ & $3.29$ \\
    $\mu^-\nu_\tau\bar\nu_\mu$ & $\pi^+\bar\nu_\tau$                                    &$2.07$ & $1.79$ & $3.92$  & $2.02$ \\
    $\mu^-\nu_\tau\bar\nu_\mu$ & $\rho^+\bar\nu_\tau$                                   &$2.03$ & $1.72$ & $3.25$  & $1.83$ \\
    $\mu^-\nu_\tau\bar\nu_\mu$ & $\mathrm{e}^+\bar\nu_\tau\nu_\mathrm{e}$               &$5.83$ & $2.28$ & $12.26$ & $3.31$ \\
    $\mu^-\nu_\tau\bar\nu_\mu$ &$\mu^+\bar\nu_\tau\nu_\mu$                              &$3.11$ & $2.32$ & $14.41$ & $3.27$
    \end{tabular}
\end{center}\end{table}

\section{Hadronic current model for tau decays}
\label{sec:hadroniccurrentmodel}
The spin-density matrices $D^\pm_{\lambda_\pm\lambda_{\pm}^\prime}$ used in Eq.~(\ref{eq:intens}) for the decays 
of the $\tau^\pm$ are constructed via:
\begin{equation}
D^\pm_{\lambda_\pm\lambda_\pm^\prime} = \mathfrak{A}_{\lambda_\pm}^{\pm*} \mathfrak{A}_{\lambda_\pm^\prime}^\pm
,\end{equation}
where $\mathfrak{A}_{\lambda_\pm}^{\pm}$ is the amplitude for the decay of a $\tau^\pm$ with helicity $\lambda^\pm$ 
into a particular final-state. For $\tau^-$ decays into hadronic final-states, this amplitude is given by:
\begin{equation}\label{eq:decampl}
\mathfrak{A}_{\lambda_-}^{-} \propto \bar u_\nu \gamma_\mu (1-\gamma^5) u_{\lambda_-}\,J_\text{had}^\mu = \ell_{\lambda_-\mu} J_\text{had}^\mu
,\end{equation}
where $J_\text{had}^\mu$ is the hadronic current describing the hadronic dynamics of the decay.
For decays into a single $\pi^-$ or $\rho^-$ and an escaping $\nu_\tau$, the corresponding hadronic currents are given by:
\begin{equation}
J_{\pi^-}^\mu \propto p_\pi^\mu\quad\text{and}\quad J_{\rho^-}^\mu \propto \text{BW}_\rho(p_\rho^2)\left(\eta^\mu_\nu - \frac{p_\rho^\mu p_{\rho\nu}}{p_\rho^2}\right) \left(p_{\pi^-}^\nu - p_{\pi^0}^\nu\right)
.\end{equation}
$\text{BW}_\rho(s)$ describes the dynamic amplitude of the intermediate $\rho(770)$ resonance,
subsequently decaying into two pions. 
Since this only acts as a scalar factor in the hadronic current, it cancels in the construction of the optimal 
observables defined in Eq.~(\ref{eq:OO}) and thus does not affect the measurement of $F_{2/3}$.

The formulation of the hadronic current 
in terms of final-state particle momenta 
for multi-body final-states\footnote{multi-body final states discussed 
here only contain three observed hadrons and do not refer to higher 
multiplicities making up $\approx30\%$ of multi-hadron decays. However,
the question of hadronic models is also present in the case of higher 
multiplicities.}
is not straightforward and requires modelling of the hadron dynamics. 
In this work, we study the decay $\tau^-\to3\pi^\pm + \nu_\tau$ and model
the hadronic current within the isobar model, following previous 
analyses \cite{cleo} and \cite{amplPaper}. In the isobar model, the total hadronic current
is composed of several {\it partial waves}, which each corresponds to a particular set 
of quantum numbers $J^{PC}$ for the
three-pion system, which subsequently decays into a $\pi^-$ and 
another known resonance finally decaying into $\pi^++\pi^-$, hereafter called the isobar.
\begin{equation}\label{eq:pwa}
J_{3\pi}^\mu = \sum_{w\in\{\text{waves}\}} c_w j_w^\mu
.\end{equation}
The complex-valued coefficients $c_w$ encode the strengths 
and relative phases of the individual partial waves, while the partial-wave 
currents $j_w^\mu$ encode their specific dependence 
on the final-state four-momenta. A detailed formulation of the $j_w^\mu$ can be 
found in Ref.~\cite{amplPaper}. Besides the isobar model presented here, there 
are other models for $J_{3\pi}^\mu$, e.g. R$\chi$T models \cite{chiral} also commonly used.

\section{Optimal observables}
\label{sec:optimalobservables}

The tau lepton form factors $F_{2/3}(q^2)$, which contain the here sought after physics observables 
$\mu_\tau$ and $d_\tau$ only enter in the description of the spin density matrix for the $\tau^{\pm}$ 
pair production [see eq.~(\ref{eq:SDMconstruction})]. We may thus single out their effect on 
the intensity $\mathcal{I}$ by rewriting equation~\ref{eq:intens}:
\begin{equation}
\mathcal{I} = \mathcal{I}_\text{SM} + \sum_{x\in\{\Re/\Im(F_{2/3})\}} x \cdot \mathcal{I}_x
,\end{equation}
where $\mathcal{I}_\text{SM}$ is the standard model intensity distribution and 
$\mathcal{I}_x$ are the specific intensity distributions corresponding the non-zero real and imaginary parts
$\Re/\Im(F_{2/3})$. Since the form factors $F_{2/3}$ are known to be small, quadratic terms in the form factors are neglected.

Each observable (form factors and thus the dipole moments) depends on specific relations among the 
measurable quantities of the final state particles. Using this expansion, we can define four optimal 
observables $OO_x$, one for each of the four $x\in\{\Re/\Im(F_{2/3})\}$, being 
optimally sensitive to the form factors \cite{OO}:
\begin{equation}\label{eq:OO}
OO_x = \frac{\mathcal{I}_x}{\mathcal{I}_\text{SM}}
.\end{equation}
Using these observables, the form factors can be 
extracted via the expectation values of the corresponding $OO_x$ obtained 
for a given data-set:
\begin{equation}\label{eq:linear}
\left\langle OO_x\right\rangle = a_x \cdot x + b_x
,\end{equation}
where the coefficients $a_x$ and $b_x$ are determined
from simulations.

\section{Studies using simulated data}
\label{sec:studies}
We now study the impact of the hadronic model on the determination 
of $F_{2/3}$ using the optimal observables defined in Sec.~\ref{sec:optimalobservables}. 
For this we construct a hadronic toy model consisting of the following nine partial waves:
\begin{equation}\label{eq:waveSet}
\begin{array}{ccc}
a_1[\rho\pi]_S&a_1[\rho\pi]_D&a_1[f_2\pi]_P\\
a_1[\sigma\pi]_P&a_1[f_0\pi]_P&\pi_1[\rho\pi]_P\\
\pi[\sigma\pi]_S&\pi[f_0\pi]_S&\pi[\rho\pi]_P
\end{array}
\end{equation}
where the naming scheme $X[\xi\pi]_L$ denotes a three-pion resonance
$X$ (the hadronic system) decaying into an isobar $\xi$ and a pion with 
relative orbital angular momentum $L$. The subsequent decay of the isobar $\xi$ into 
two pions is implied and in turn described by a set of known decay amplitudes.
 Each resonance $X$ represents a set of quantum numbers $J^{PC}$.

For the model, we used partial-wave coefficients $c_w$ loosely inspired 
by a partial-wave analysis of the three-pion final state in Ref.~\cite{compass}. The dominant 
wave in this model is the $a_1[\rho\pi]_S$ wave, as is expected following previous 
analyses \cite{cleo}. Using our toy model,
we generated data sets with $10^6$ $\tau^\pm$-pair events, where the $\tau^-$ 
decays into $(3\pi^\pm + \nu_\tau)$ according to the model described above, 
while the $\tau^+$ decays into $(\pi^++\bar\nu_\tau)$. In total, we generated 
four toy data sets, where one of each of the four $\Re/\Im(F_{2/3})$ takes the 
value of 0.01, while the other three values remain 0.

In a first study, we analyze the pseudo data using the same hadronic 
model as used for the simulation
and extract the form-factors. We found no 
bias and an accuracy comparable to the other hadronic decay modes $(\pi^- + \nu_\tau)$
and $(\rho^- + \nu_\tau)$ for the same number of events is obtained. For $10^6$ simulated events, 
we find:
\begin{equation}
\begin{array}{ll}
\delta\Re(F_2) = 0.0006;&\delta\Im(F_2) = 0.0007;\\
\delta\Re(F_3) = 0.0009;& \delta\Im(F_3) = 0.0005,
\end{array}
\end{equation}
In a second study, we analyzed the same simulated data sets but now using 
a simplified model for the hadronic current, namely now only comprising the $a_1[\rho\pi]_S$
wave. 
To quantify the similarity of two hadroic models, we define the model overlap $\omega_{m,m^\prime}$ 
of two models $m$ and $m^\prime$ for the hadronic current as the normalized product of the total hadronic 
currents $J^\mu_m$, contracted with the corresponding leptonic current $\ell_{\lambda_-}^\mu$ and integrated 
over the full Lorentz invariant phase space (LIPS):\footnote{The 
overlaps are the same for $\lambda_-=\pm1/2$. The normalization factors $\mathcal{N}_m$ 
ensure $\omega_{m,m} = 100\%$.}:
\begin{equation}
\omega_{m,m^\prime} = \left|\int\mathrm{dLIPS}\,\left( J_m^{\mu}\ell_{\lambda_-\mu}\right)^*\left(\ell_{\lambda_-\nu}J_{m^\prime}^{\nu}\right)\right| 
\Bigg/ \Big(\mathcal{N}_m\cdot\mathcal{N}_{m^\prime}\Big)
,\end{equation}
with the leptonic current $\ell_{\lambda_-}^\mu$ defined in Eq.~(\ref{eq:decampl}).
The model-overlap of $\omega_{\text{true},\text{ana}}$ of the simplified model 
with the model used for simulation was $78\%$.

For this study, we also re-determined the coefficients $a_x$ and $b_x$ defined in Eq.~(\ref{eq:linear}) 
so that they correspond to our simplified analysis model.
Repeating our analysis with a wrong hadronic model results in the 
following values for $\Re/\Im(F_{2/3})$:
\begin{equation}
\begin{array}{ll}
\Re(F_2) = 0.0529\pm 0.0008;&\Im(F_2) = 0.0118\pm0.0008;\\
\Re(F_3) = 0.0086\pm0.0012;&\Im(F_3) = 0.0079 \pm 0.0005,
\end{array}
\end{equation}
while the true value for these quantities is always 0.01. We find, that 
$\Re(F_2)$ is largely over-estimated, while the effect in $\Im(F_2)$ is not 
very large. $\Re(F_3)$ and $\Im(F_3)$ suffer an under-estimation, which, 
however, is less than for $\Re(F_2)$. If the true value is set to $0$, 
the bias in $\Re(F_2)$ persists, while we observe no bias for $\Re/\Im(F_3)$
in this case.

We now repeated this procedure with different de-tuned analysis 
models for every individual partial wave given in Eq.~(\ref{eq:waveSet}). 
For this, we scale up one individual partial wave coefficient $c_w$ [see Eq.~(\ref{eq:pwa})]
from the true model such, that the model overlap $\omega_{\text{true},\text{ana}}$ 
drops to $95\%$, while keeping the remaining coefficients at their nominal values. 
Doing so, we find that the values obtained for $F_3$ and $\Im(F_2)$ are consistent 
with the input values, regardless of the wave scaled.
Thus, the extraction of these three quantities appears to be 
rather robust with respect to changes in the hadronic model.

In the case of $\Re(F_2)$, we observe a significant bias due to the 
mismatch between generator and analysis hadronic model. This bias depends on the 
individual partial wave that is scaled in the particular study and is given 
in Tab.~\ref{tab:reFtwoBias}.

\begin{table}\begin{center}
\caption{$\Re(F_2)$ extracted from a simulated data set with an 
input value of $\Re(F_2) = 0.01$, analyzed with a single de-tuned partial wave.
The statistical uncertainties of all values shown are $0.0007$.}
\begin{tabular}{l|lllll}\label{tab:reFtwoBias}
De-tuned wave & $a_1[\rho\pi]_S$&
$a_1[\rho\pi]_D$&$a_1[f_2\pi]_P$ &$a_1[\sigma\pi]_P$&$a_1[f_0\pi]_P$\\
$\Re(F_2)$ &0.0178&0.0168&0.0144&0.0169&0.0143\\\hline
De-tuned partial wave & $\pi_1[\rho\pi]_P$ & $\pi[\sigma\pi]_S$&$\pi[f_0\pi]_S$&$\pi[\rho\pi]_P$ &\\
$\Re(F_2)$ &0.0147&0.0186&0.0162&0.0180&
\end{tabular}
\end{center}\end{table}

In a final study, we de-tuned the $a_1[\rho\pi]_S$-wave 
such that the model-overlap $\omega_{\text{true},\text{ana}} = 99\%$.
In this case, we obtain:
\begin{equation}
\begin{array}{ll}
\Re(F_2) = 0.0112\pm 0.0007;&\Im(F_2) = 0.0102\pm0.0007;\\
\Re(F_3) = 0.0097\pm0.0009;&\Im(F_3) = 0.0103 \pm 0.0005.
\end{array}
\end{equation}
Thus, we find that a proper model for the hadronic current $J_{3\pi}^\mu$
alleviates possible bias in the determination of $\Im(F_2)$ and $F_{3}$, while 
the bias in $\Re(F_2)$ remains significantly larger than the uncertainty, 
even for a model overlap very close
to unity. Since $\Re(F_2)$ is the only quantity that alters the total cross-section 
(see Fig.~\ref{fig:SDM}), it might be advisable to neglect the spin-information 
of the decays and only use the total $\tau^\pm$-pair production cross-section. 
Doing so, we find for the same simulated data introduced above:
\begin{equation}\label{eq:crossSectionResolution}
\Re(F_2) = 0.0108 \pm 0.0015
.\end{equation}
Even though the accuracy for $\Re(F_2)$ is worse by a factor of two, this result is independent 
of a hadronic model and thus is not affected by model bias. 
Including only the spin-information from the $(\pi^++\bar\nu_\tau)$ decay does not 
improve the accuracy given in Eq.~(\ref{eq:crossSectionResolution}). This is expected, 
since $\Re(F_2)$ only affects the correlation of both $\tau^\pm$ spins.
However, the measurement of the total cross-section requires that all radiative corrections 
are known and is typically very difficult, since it introduces new sources of systematic uncertainties.

Evaluating the distributions from Fig.~\ref{fig:SDM} for each partial wave, we could not single 
out particular waves being specifically more sensitive to the observation of EDM or MDMs than others. 
The scheme of optimized variables would, however, take into account such possible effects.

\color{black}
\section{Conclusion}
\label{sec:conclusions}
We studied the determination of the 
tauon form factors $F_2$ and $F_3$ using simulated $(3\pi^\pm + \nu_\tau)$$\times$ $(\pi^++\bar\nu_\tau)$ 
$\tau^\pm$-events. We find the $3\pi^\pm$ hadronic final-state 
to give an accuracy on the form-factors comparable to 
other hadronic channels, assuming the model 
for the hadronic current $J_{3\pi}^\mu$ to be perfect. Thus, this 
decay channel will help to significantly increase usable 
data for purely hadronically decaying $\tau^\pm$-pair events.
For a simulated data set of $10^6$ events, we obtain an 
accuracy for $\mu_\tau$ and $d_\tau$ of:

\begin{equation}
\begin{array}{ll}
\delta\Re(\mu_\tau) = 3.46\times10^{-18}e\text{cm}& \delta\Im(\mu_\tau) =3.58\times10^{-18}e\text{cm}\\
\delta\Re(d_\tau) = 4.66\times10^{-18}  e\text{cm}& \delta\Im(d_\tau) = 2.61\times10^{-18}e\text{cm}.
\end{array}
\end{equation}
However, the model for $J_{3\pi}^\mu$ is not known a priori 
and all models currently used, e.g. the isobar model or
R$\chi$T models \cite{cleo,amplPaper,chiral}, are based on assumptions,
a perfect hadronic model is currently not available.
Thus, we extended our study to hadronic models for $J^\mu_{3\pi}$ that 
differ from the true model and found a small bias in the 
extraction of $F_3$ and $\Im(F_2)$, while $\Re(F_2)$ is heavily over-estimated.

The observed bias results in an under-estimation of $\Re/\Im(F_3)$,
which in turn vanishes as the analysis model approached 
the true model. The bias in $\Re(F_2)$, however, remains significant 
even at a model overlap $\omega_{\text{true},\text{ana}}=99\%$ and 
thus seems to prohibit the use of the $3\pi^\pm$ channel in a determination 
of $\Re(F_2)$. However, since $\Re(F_2)$ alters the total $\tau^\pm$ pair 
production cross-section, we may ignore spin effects for such final states and still determine 
$\Re(F_2)$. Ignoring spin-correlations decreases the accuracy by 
a factor of two, but removes the strong model-dependence.

Finally, we stress that a good knowledge of the hadron 
dynamics of multi-particle $\tau^\pm$ decays is prerogative for their inclusion 
in precision measurements like $F_{2/3}(q^2)$.
A simple approximation of the hadronic current by the dominating 
$a_1\to[\rho\pi]_S$ contribution does not suffice, since according to current knowledge it only describes 
around $70\%$ of the $\tau\to3\pi+\nu$ intensity \cite{cleo}.

%
%
%
%



\bibliography{SciPost_Example_BiBTeX_File.bib}

\nolinenumbers

\end{document}